\documentclass[12pt,a4paper]{article}
\usepackage{amsmath,amssymb}
\usepackage{graphicx,epsfig}
\usepackage{color}

\newcommand{\lsim}{\, \mathop{}_{\textstyle \sim}^{\textstyle <} \,}

\begin{document}

\begin{titlepage}
\begin{flushright}
UCB-PTH-05/34\\
LBNL-59009
\end{flushright}

\vskip 1.0cm

\begin{center}
  {\Large \bf Improved Naturalness and the Two Higgs Doublet Model}
 
   \vskip 1.0cm
 
  {\large Riccardo Barbieri$^a$  and  Lawrence J.~Hall$^b$}\\[1cm]
{\it $^a$ Scuola Normale Superiore and INFN, Piazza dei Cavalieri 7, I-56126 Pisa, Italy} \\[5mm]
{\it $^b$ Department of Physics, University of California, Berkeley, and\\
  Theoretical Physics Group, LBNL, Berkeley, CA 94720, USA}\\[5mm]

     \vskip 1.0cm
 
   \abstract{The natural cutoff scale for the quadratically divergent 
   top quark contribution to the Higgs mass parameter can be significantly
   raised above
   the surprisingly low standard model value, with important consequences
   for the  LHC:  the physics that cancels the top quark divergence may 
   be out of reach, while an electroweak sector with ``improved naturalness''
   may be discovered. Such a sector, consistent with electroweak
   precision tests,  arises
   in the two Higgs doublet model with heavy Higgs and top quark interactions 
   that approach strong coupling.}

\end{center}
\end{titlepage}
\def\baselinestretch{1.05}

\section{Improved Naturalness}

The Standard Model (SM) is inadequate as a fundamental theory because
quadratic divergences in the Higgs mass parameter lead to a high
sensitivity of electroweak physics to large energy scales.  The dominant
quadratic divergence arises from virtual top quarks, requiring new physics
to appear at or below the scale
\begin{equation}
\Lambda_t \lesssim 400 \; \mbox{GeV} \left( {\frac{m_h}{115 \; \mbox{GeV}}}
\right){D_t^{1/2}},
\label{eq:LSM}
\end{equation}
where $D_t$ is the sensitivity  of the Higgs mass to $\Lambda_t$; 
or equivalently,  the amount of fine tuning necessary to 
make $\Lambda_t$ large is 1 in $D_t$. Since electroweak precision
tests (EWPT) indicate that the Higgs boson is light,
$m_h < m_{EW} = 285 \mbox{GeV}$ at 95\% CL \cite{EWPT} with a central
value close to the lower bound of 115 GeV from direct searches, the
new physics at $\Lambda_t$ should be accessible to the LHC.  What is
this new physics? The most ambitious attempt at a complete theory
incorporates supersymmetry at the weak scale to understand
the hierarchy problem, with the top squark
cancelling the quadratic divergence of the top quark.  
While the simplest model requires some fine 
tuning, supersymmetry remains, perhaps, the leading candidate.  
A less ambitious approach is
to delay the need for fine tuning up to 5---10 TeV, yielding a
little hierarchy that at least explains why higher dimension operators
from this cutoff scale do not upset the success of a light SM Higgs
with EWPT. Such theories have their own answer for the new physics
that cancels the top quark quadratic divergence; for example, in Little
Higgs models the LHC will discover new vector-like quarks. The
simplest theories, however, require some 
amount of fine tuning to agree with EWPT\footnote{For a recent analysis see Ref. \cite{Marandella:2005wd}  and references therein.}.

In this letter we pursue an alternative idea, of limited scope, that
nevertheless has crucial implications for the LHC. Is it possible to
construct theories where $\Lambda_t$ is modestly increased above
(\ref{eq:LSM}), for example by a factor of 3---5, so that the new
physics associated with this scale is inaccessible to LHC?  If so,
what is the effective field theory below $\Lambda_t$---certainly not
the SM---how is it consistent with EWPT, and what signals does it
give at the LHC?  While the ``LEP paradox''\cite{Barbieri:2000gf}
may not be solved, it is nevertheless ameliorated, and the implications for the LHC
are immediate. Instead of focussing on the new physics that cancels
the top quark divergence (squarks, vector quarks, ...) one must study
the physics of the modified electroweak theory below $\Lambda_t$.  
In a previous paper it was shown that such a theory can result from
mixing the Higgs with a Higgs of a mirror world \cite{BGH, Chacko:2005pe}. 
In this paper we demonstrate that Higgs mixing in the two Higgs doublet 
model provides a conceptually simpler example.

%%%%%%%%%%%%%%%%%%%%%%%%%%%%%%%%%%%%%%%%%%%%%%%%%%%%%%%%%%%%%%
\section{The Two Higgs Doublet Model }

The scalar potential for the most general two Higgs doublet model that
satisfies natural flavor conservation \cite{Glashow:1976nt} is \cite{Gunion:1989we}
\begin{eqnarray}
V & = & 
- \mu_1^2 H_1^\dagger H_1 - \mu_2^2  H_2^\dagger H_2 
+ \lambda_1 (H_1^\dagger H_1)^2 + \lambda_2 (H_2^\dagger H_2)^2
+ \lambda_3 H_1^\dagger H_1 H_2^\dagger H_2 \\
&+& \lambda_4 H_1^\dagger H_2 H_2^\dagger H_1
+ \lambda_5 [(H_1^\dagger H_2)^2 + (H_2^\dagger H_1)^2]. 
\label{eq:V}
\end{eqnarray}
A discrete symmetry acts on $H_2$ so that it alone couples to the up
quarks---in particular to the top quark---while the down quark masses
can arise from either an interaction with $H_1$ or $H_2$. In a phase
where both Higgs doublets acquire vevs, $H_i = (0, v_i + h_i)$ with
$v_{1,2} \neq 0$ and real, the mass matrix for the two neutral 
Higgs bosons is \cite{Gunion:2002zf}
\begin{equation}
V_2 = \left( h_1, h_2 \right)
\left( \begin{array}{cc} 4 \lambda_1 \; v_1^2 &  2 \delta \; v_1 v_2  \\
                         2 \delta \; v_1 v_2 &  4 \lambda_2 \; v_2^2 
                         \end{array} \right)
\left( \begin{array}{c} h_1 \\ h_2 \end{array} \right),
\label{eq:V2}
\end{equation}
where $\delta = \lambda_3 + \lambda_4 + 2 \lambda_5$. Motivated by the
possibility of raising  $\Lambda_t$ in a way consistent with EWPT, we
assume that the 22 entry is the largest. Thus the heaviest Higgs boson is 
$h_+ = \cos \alpha \; h_2 + \sin \alpha \; h_1$, with mass
\begin{equation}
m_+^2 \approx 4 \lambda_2 \; v_2^2,
\label{eq:m+}
\end{equation}
and the Higgs mixing angle is small
\begin{equation}
\alpha \approx  \frac{\delta}{2 \lambda_2} \; \frac{v_1}{v_2}.
\label{eq:alpha}
\end{equation}
This contrasts with the minimal supersymmetric standard model, where
it is the lightest Higgs boson that couples dominantly to the top
quark. Of the seven parameters of the potential, five $(\mu_1^2,
\mu_2^2, \lambda_1, \lambda_2, \delta)$ can be specified as $v =
\sqrt{v_1^2 + v_2^2}, \;  \tan \beta = v_2/v_1, \; m_\pm^2$ and $\alpha$.
The remaining two parameters $\lambda_{4,5}$ can be traded for the
charged scalar mass $m_{H^-}$ and the pseudoscalar mass $m_A$.

\section{The Scale $\Lambda_t$ and Electroweak Precision Tests}
%%%%%%%%%%%%%%%%%%%%%%%%%%%%%%%%%%%%%%%%%%%%%%%%%%%%%%%%%%%%%%%

The quadratic divergence induced at 1 loop
by virtual top quarks appears only in the parameter $\mu_2^2$
and is cutoff at some scale $\Lambda_t$
\begin{equation}
\mu_2^2 = \mu_0^2 + a_t \Lambda_t^2,
\label{eq:mu2}
\end{equation}
where $\mu_0^2$ is the bare parameter,  $a_t = 3 \lambda_t^2/8 \pi^2$
and $\lambda_t = m_t/(v \sin \beta)$.
The sensitivity of the Higgs masses, $m_\pm^2$, to the scale $\Lambda_t$ is given by
$D_t^\pm \equiv \partial \ln m_\pm^2 / \partial \ln \Lambda_t^2$,
which can be inverted to give
\begin{equation}
\Lambda_t 
= \left( \frac{2 \pi v}{\sqrt{3}} \frac{m_+}{m_t} \right) 
\left\{ \begin{array}{c} \sin \beta \sqrt{D_t^+} \\
                        \cos \beta \sqrt{D_t^-} \sqrt{2
			  \lambda_1/\delta}
\end{array} \right. .
\label{eq:Lambdat}
\end{equation}
The usual SM result (\ref{eq:LSM}) is given by the first parenthesis
with the replacement $m_+ \rightarrow m_h$.  In the SM $m_h$ is limited
by EWPT, so that a crucial question becomes how EWPT limit $m_+$. 
%We assume that $\delta$ is not significantly larger than $2 \lambda_1 $. 
%This implies $m_-^2 \approx 4 \lambda_1 v_1^2$ 
%but does not directly constrain $m_{H^-}$ and $m_A$.
We assume that neither $\sin \beta$ nor $\cos \beta \sqrt{2
\lambda_1/\delta}$ are small, so that a substantial increase in
$m_+$ above $m_h$ guarantees an increase in $\Lambda_t$ above the SM
value. 

The quantity  $\cos \beta \sqrt{2 \lambda_1/\delta}$ can be
rewritten as $m_- / \sqrt{2 \lambda_3 v^2 - m_{H^-}^2}$. If $m_{H^-}$
is relatively light, for example 200 GeV, this factor can be naturally of order
unity. Such a light charged Higgs contributes radiative corrections to
the $Z \bar{b} b$ vertex $g_L$,  contributing an amount to $R_b$ that, for
$\tan \beta = 1$, corresponds to about 1 standard deviation in
the measured value. While formally this can be used to place limits on
$m_{H^-}$ \cite{Haber:1999zh}, such limits may prove untrustworthy as 
$A_b$ lies nearly three standard deviations from the SM 
prediction \cite{EWPT}. 
For example, if some new physics provides a contribution to $g_R$ to
account for $A_b$, then a contribution to $g_L$ from a light charged Higgs
would be welcome. Nevertheless, this radiative correction to $g_L$ is
proportional to $\cot^2 \beta$, so that $v_2$ should not be much less than
$v_1$. For $\tan \beta$ close to 1,
there is a limit of about 200---250 GeV on $m_{H^-}$ from $b
\rightarrow s \gamma$ \cite{neubert, Gambino:2001ew}. Even if the charged Higgs boson
is much heavier than this, a large $\Lambda_t$ is still possible
provided there is a cancellation between $m_{H^-}^2$ and $2 \lambda_3
v^2$. 

Do EWPT allow large values for $m_+$?
In the two Higgs doublet model with $m_{H^-} = m_A$, the contributions
of the scalars to the $S$ and $T$ parameters take the 
form\footnote{The  explicit expression for the radiative corrections 
in a general two Higgs doublet model can be found in Ref. \cite{Hagiwara:1994pw}.}
\begin{equation}
S = cos^2 \beta \; S_{SM}(m_-) +  sin^2 \beta \; S_{SM}(m_+) + \Delta S 
\label{eq:S}
\end{equation}
\begin{equation}
T = cos^2 \beta \; T_{SM}(m_-) +  sin^2 \beta \; T_{SM}(m_+),
\label{eq:T}
\end{equation}
where $S_{SM}(m_h)$ and  $T_{SM}(m_h)$ are the contributions in the
SM, and we have approximated $\beta - \alpha$ by $\beta$
because $\alpha$ is small. Consider first the approximation that  
$S_{SM}(m_h)$ and $T_{SM}(m_h)$ both have the form $A + B \ln m_h$ and
that $\Delta S$ can be neglected. In this case the scalar contributions to
$S$ and $T$ in the two Higgs doublet model are obtained from those in
the SM by the replacement
$\log{m_h} \rightarrow \cos^2 \beta \log{m_-} + \sin^2 \beta \log{m_+}$
implying that the current 95 \% C.L. limit on the SM Higgs mass, 
$m_h < m_{EW} \approx 285$ GeV, gets replaced by
\begin{equation}
m_-^{\cos^2 \beta} m_+^{\sin^2 \beta} < m_{EW}
~~~~~ \mbox{or} ~~~~~
m_+ < m_- \left( \frac{m_{EW}}{m_-} 
\right)^{\frac{1}{\sin^2 \beta}}.
\label{eq:EWPT}
\end{equation}
This shows that for $m_- < m_{EW}$ the bound on $m_+$ gets
exponentially relaxed as $\sin \beta$ is reduced. Thus we are led to
consider low values for $m_-$ and $v_2 \lsim v_1$. The factor
of $\sin \beta$ in the upper line of (\ref{eq:Lambdat}) is sub-dominant to the
exponential behaviour of $m_+$. The point is very simple: as $v_2$ is
reduced so the state $h_+$, which is mainly $h_2$, decouples from
electroweak breaking phenomena. Furthermore, because the bound on $m_+$ has an
exponential dependence on $\sin \beta$, $v_2$ need not be much less
than $v_1$. Indeed, $v_2$ cannot be reduced too much as we have
assumed that $4 \lambda_2 v_2^2$ is the largest term in the Higgs
boson mass matrix, so that reducing $v_2$ leads to strong coupling in
$\lambda_2$. The perturbativity limit on $m_+$ is $4 \pi v \; \sin
\beta$. 
For $m_-$ close to the direct search limit of 115 GeV, a value
of $\sin \beta =$ 0.6---0.7 is
sufficient to raise $m_+$ to near a TeV. The cutoff scale $\Lambda_t$
is raised above the SM value by a factor of 5 to 2 TeV.

In the above analysis we have ignored $\Delta S$, which is not
justified since it involves a large logarithm as $m_+$ is made
large. We find that at 1 loop order
\begin{equation}
\Delta S = \frac{\cos^2 \beta}{6 \pi} \left( \ln \frac{m_+}{m_A} 
  \right) 
\label{eq:DeltaS}
\end{equation}
for $m_+ \geq m_{H^-} = m_A \geq m_-$, up to a small non-logarithmic term .  
This positive contribution to $S$
gives a final result that is well within the 90\% C.L. region of fits to electroweak 
data.  The reason why this data excludes the SM for large $m_h$
while allowing the two Higgs doublet model with large $m_+$ is that
the large logarithm making a negative contribution to $T$ in the SM
has a coefficient reduced by $\sin^2 \beta$ in the two Higgs doublet theory.

\section{Summary}

The leading question of electroweak symmetry breaking is often taken
to be: what is the physics that cancels the quadratic divergence of
the top quark, and what signals does it have at the LHC?  We have
pointed out that this might not be the right question as far as LHC
physics is concerned.  A theory with improved naturalness compared 
to the SM may place the physics of top-cancellation beyond the reach
of the LHC, while leaving a non-SM Higgs sector to explore. We have
shown that the two Higgs doublet model is able to accomplish this,
with a light Higgs $m_- \approx$ 115---150 GeV, a heavy Higgs $m_+
\approx $ 500 GeV---1,000 GeV, relatively small Higgs mixing $\alpha \lesssim$
1/3  and $\tan \beta \approx$ 0.8---1. These ranges are only
meant as a rough guide, but the underlying physics is that one should
think of the theory as having two sectors: the $(t, h_2)$ sector which
is approaching strong coupling, and the perturbative $(W, h_1)$ sector
with slightly more of the $W$ mass coming from $v_1$ than $v_2$.

\section{Acknowledgments}

This work is supported by the EU under RTN contract
MRTN-CT-2004-503369. The work of R.Barbieri is also supported by MIUR.
The work of L.J.Hall was supported by the US Department of Energy under
Contract DE-AC03-76SF00098 and DE-FG03-91ER-40676, and by the
National Science Foundation under grant PHY-00-98840.


\begin{thebibliography}{99}
  
  \bibitem{EWPT}
  The ALEPH, DELPHI, L3, OPAL and SLD Collaborations
  and the LEP Electroweak Working Group, the SLD Electroweak and Heavy Flavour Groups, 
  [arXiv:hep-ex/0509008].

%\cite{Marandella:2005wd}
\bibitem{Marandella:2005wd}
  G.~Marandella, C.~Schappacher and A.~Strumia,
  %``Little-Higgs corrections to precision data after LEP2,''
  Phys.\ Rev.\ D {\bf 72}, 035014 (2005)
  [arXiv:hep-ph/0502096].
  %%CITATION = HEP-PH 0502096;%%
  
%\cite{Barbieri:2000gf}
\bibitem{Barbieri:2000gf}
  R.~Barbieri and A.~Strumia,
  %``The 'LEP paradox',''
  [arXiv:hep-ph/0007265].
  %%CITATION = HEP-PH 0007265;%%
  
  
\bibitem{BGH}
R. Barbieri, T. Gregoire and L.J. Hall, [arXiv:hep-ph/0509242].
  %\cite{Chacko:2005pe}
\bibitem{Chacko:2005pe}
  Z.~Chacko, H.~S.~Goh and R.~Harnik,
  %``The twin Higgs: Natural electroweak breaking from mirror symmetry,''
  [arXiv:hep-ph/0506256].
  %%CITATION = HEP-PH 0506256;%%

  
  %\cite{Glashow:1976nt}
\bibitem{Glashow:1976nt}
  S.~L.~Glashow and S.~Weinberg,
  %``Natural Conservation Laws For Neutral Currents,''
  Phys.\ Rev.\ D {\bf 15}, 1958 (1977).
  %%CITATION = PHRVA,D15,1958;%%
  
  %\cite{Gunion:1989we}
\bibitem{Gunion:1989we}
  J.~F.~Gunion, H.~E.~Haber, G.~L.~Kane and S.~Dawson,
  %``The Higgs Hunter's Guide,''
SCIPP-89/13.

%\cite{Gunion:2002zf}
\bibitem{Gunion:2002zf}
  J.~F.~Gunion and H.~E.~Haber,
  %``The CP-conserving two-Higgs-doublet model: The approach to the  decoupling
  %limit,''
  Phys.\ Rev.\ D {\bf 67}, 075019 (2003)
  [arXiv:hep-ph/0207010].
  %%CITATION = HEP-PH 0207010;%%
  
\bibitem{Haber:1999zh}
  H.~E.~Haber and H.~E.~Logan,
  %``Radiative corrections to the Z b anti-b vertex and constraints on  extended
  %Higgs sectors,''
  Phys.\ Rev.\ D {\bf 62}, 015011 (2000)
  [arXiv:hep-ph/9909335].
  %%CITATION = HEP-PH 9909335;%%

\bibitem{neubert}
M. Neubert,
Eur.Phys.J. C40 (2005) 165-186,
[arXiv:hep-ph/0408179].

%\cite{Gambino:2001ew}
\bibitem{Gambino:2001ew}
  P.~Gambino and M.~Misiak,
  %``Quark mass effects in anti-B $\to$ X/s gamma,''
  Nucl.\ Phys.\ B {\bf 611}, 338 (2001)
  [arXiv:hep-ph/0104034].
  %%CITATION = HEP-PH 0104034;%%
  

  %\cite{Hagiwara:1994pw}
\bibitem{Hagiwara:1994pw}
  K.~Hagiwara, S.~Matsumoto, D.~Haidt and C.~S.~Kim,
  %``A Novel approach to confront electroweak data and theory,''
  Z.\ Phys.\ C {\bf 64}, 559 (1994)
  [Erratum-ibid.\ C {\bf 68}, 352 (1995)]
  [arXiv:hep-ph/9409380].
  %%CITATION = HEP-PH 9409380;%%
  
\end{thebibliography}
\end{document}